# Role of cavitation in drying cementitious materials

Monisha Rastogi[ab], Arnaud Müller[a], Mohsen Ben Haha[a], Karen L. Scrivener[b]

[a]HeidelbergCement Technology Center, Leimen, Germany

[b]École polytechnique fédérale de Lausanne, Switzerland

*Abstract*

This study aims to identify and isolate the role of properties of water with respect to the shape of the first desorption isotherm of hardened cement pastes. It has been observed that desorption occurs not only by receding meniscus but is also affected by cavitation. Classical nucleation theory was used to understand and examine parameters that influence cavitation. The results show that the sudden moisture loss at 0.3 $p_v/p_{sat}$ in desorption isotherm occurs due to homogeneous cavitation. Further investigations revealed that homogeneous cavitation occurs in the part of gel porosity that is disconnected from the rest of capillary pores in the pore network.

# 1. Introduction

Cementitious materials have excess water in their porous structure which is left after the hydration process. The difference in the relative humidity between the internal surface of cementitious material and exposed environment could result in evaporation (or desorption) of this water or absorption of the water vapor from the external environment into their porous structure. The changes in moisture content in the cementitious materials lead to dimensional changes (swelling or shrinkage) and cause several other degradation processes.

Moisture sorption isotherms can be useful in understanding the moisture change response of cementitious materials and therefore help with durability assessments.[1–4]. The first desorption isotherm is particularly relevant as it gives information on as-cured samples.

A step or a sudden jump in the moisture loss content around 0.3 $p_v/p_{sat}$ [1] or 30 % relative humidity (RH) is usually observed in the desorption isotherm of cementitious materials [5–8]. For instance, Baroghel-Bouny investigated several cement pastes (with and without incorporation of silica fume) and concretes with the broad range of water-to-cement ratio ranging from 0.20 to 0.84. This step around 0.3 $p_v/p_{sat}$ was observed for all the samples [9]. From sorption data, pore size distribution can be determined. When Barrett, Joyner, and Halenda (BJH) method was applied for obtaining pore size distribution, the sudden moisture loss at 0.3 $p_v/p_{sat}$ was transposed into a large population of pores of radius around 1.5 nm which was attributed to an actual pore mode (or threshold pore entry radius) in the investigations made by Baroghel-Bouny as well as several other studies [9–12]. Nevertheless, such a high concentration of pores of this radius has not been observed by any other technique such as $^1$H NMR relaxometry.

---

[1] Symbols $p_v$ and $p_{sat}$ denote the pressure imposed by the vapor and the saturation vapor pressure of water at the given temperature, respectively

The sudden moisture loss or step at 0.3 $p_v/p_{sat}$ has also been explained using several other arguments such as microstructural changes, phase-specific dehydration of $Ca^{2+}$ ions or an associated change in basal space of calcium-silicate-hydrate (C-S-H) [9][13,14]. These suppositions, however, fail to explain the occurrence of this step at the same location in the desorption isotherm irrespective of the difference in the composition of raw materials, water to binder ratio or synthesis procedure. Since the removal of the water is the common factor to all these studies, therefore the role of water towards this step should also be investigated.

Water (or electrolyte containing dissolved ions) filled pores in cementitious materials can be divided into four pore categories according to $^1$H NMR relaxometry: (I) Interlayer spaces corresponding to the water present in spaces between the solid C-S-H sheets. The characteristic size of such spaces is reported to be around 1 nm. (II) Gel pores, which are usually viewed as spaces between agglomerated C-S-H sheets with characteristic size between 2-8 nm. (III) Interhydrate pores that are the pores between C-S-H needles with a characteristic size of 8-10 nm. (IV) Capillary pores that are the larger pores containing unreacted water or air with a characteristic size spanning from 10 nm to micrometre size [15,16].

Due to the confinement of water in such an interconnected porous network, evaporation is not only governed by receding (or withdrawing) meniscus alone but can be significantly affected by cavitation. Cavitation corresponds to the drastic change of the liquid phase into vapor phase resulting from homogeneous or heterogeneous nucleation. Evaporation induced cavitation has been reported to occur in several domains of biology, materials science, and engineering such as in mineral inclusions, natural and synthetic trees, propagation of spores, lipid bilayers, and cell membranes among others [17]. Recently, Maruyama *et al.* studied the role of cavitation in hardened cement paste during short-term water vapor desorption. The authors investigated several porous materials including zeolites, portlandite, and MCM-41. The high moisture loss observed around 0.3 $p_v/p_{sat}$ in the desorption isotherm of these materials was attributed to cavitation. Furthermore, the authors reported the influence of dissolved ions and surface effects such as hydrophobicity/hydrophilicity on the cavitation pressure [18].

This paper aims to show and explain cavitation in the desorption isotherm of hardened cement pastes, using classical nucleation theory (CNT). According to CNT, surface tension[2] at the liquid-vapor interface was identified to be an important parameter that affects cavitation pressure. To change surface tension, the sorption temperature was increased. Furthermore, samples were synthesized with surface reducing admixture (SRA) to modify the cavitation pressure as SRA also reduces the surface tension at liquid-vapour interface.

## 2. Theory

It is useful to briefly review the theories, concepts and equations that are used to analyse and explain the results of this study.

### 2.1 The Barret-Joyner-Halenda (BJH) method

BJH method was initially proposed to determine the pore size distribution in the mesopore range (pore size between 2 and 50 nm, according to IUPAC classification) and in the micropore range (< 2 nm) from gas desorption isotherms[19]. This method is also extensively used to obtain pores size distribution from moisture sorption isotherms of cementitious materials [9,10,12]. BJH method considers the desorption isotherm as a series of steps and divides the pore sizes into groups such that all the pores in each group have an average radius $r_P$. As the pressure is reduced, the amount of adsorbate is removed from the inner core of the pore ($r_k$) and the thickness of the adsorbed film is reduced ($\Delta t_c$) in the first group of pores. Further decrease in pressure not only removes adsorbate from the core of second group of pore sizes, but also decreases the thickness of adsorbed layer for the first group of pores. For each group the pore core and the change in the thickness is calculated as $p_v/p_{sat}$ is lowered. The total pore volume and the surface area of the pores are obtained by adding the values in each group of pores.

---

[2] In this study, bulk values of surface tension are used without Tolman type corrections for the curvature dependence as the objective is to show the qualitative mechanism of cavitation.

## 2.2 Kelvin and Young-Laplace equation

In this study, Kelvin and Young-Laplace equation has been used to calculate the intrinsic pore capillary pressure as the function of $p_v/p_{sat}$.

Upon exposing the sample to sub-saturated vapour ($p_v/p_{sat} < 1$), the curvature of the meniscus increases and tensile stress is induced in the pore fluid. Thermodynamic equilibrium is achieved if the chemical potential of water in the pores is equal to that of the imposed vapour. If vapor is an ideal gas and the liquid is incompressible, the relation between intrinsic pore pressure and imposed vapor pressure can be given by Kelvin equation as:

$$P = p_{sat} + \frac{RT}{v_m} \ln\left(\frac{p_v}{p_{sat}}\right) \tag{1}$$

$P$ : intrinsic pore capillary pressure

$v_m$: molar volume of water (1.805 × 10⁻⁵ m³/mol)

R : Universal gas constant (8.314 J/K)

T: Temperature (in Kelvin)

Further, from the Young–Laplace equation,

$$P - p_v = -\frac{2\sigma_{lv}\cos\theta_r}{r_k} \tag{2}$$

$\sigma_{lv}$ : surface tension at the liquid-vapour interface,

$\theta_r$ : contact angle of the meniscus in a pore of radius $r_k$.

The original Young-Laplace equation (equation 2) does not consider any fluid-wall interactions. Therefore, to account for these interactions the statistical thickness of the adsorbed layer before evaporation, $t_c$ is added to the core radius, $r_k$. The corrected pore radius $r_P$ for a cylindrical pore, is given by

$$r_P = r_k + t_c \qquad (3)$$

This statistical thickness of the adsorbed layer as the function of $p_v/p_{sat}$ can be obtained by several empirical and theoretical methods which include Halsey's equation, Hagymassy t -curve, Badmann equation, from Brunauer- Emmett-Teller (BET) surface area, among others[20,21]. In this study, BET surface area is used to calculate statistical thickness.

Considering that intrinsic pore pressure $P$ is much larger in magnitude (≈ MPa) than $p_v$ and $p_{sat}$ (≈ KPa) and combining equations (1) (2) and (3), we obtain the modified Kelvin-Laplace equation which is given as follows:

$$r_P = -\frac{2\sigma_{lv}\cos\theta_r}{\frac{RT}{v_m}\ln\left(\frac{p_v}{p_{sat}}\right)} + t_c \qquad (4)$$

Assuming the contact angle as zero, as $p_v/p_{sat}$ decreases, the evaporation will occur progressively with decreasing radii.

### 2.3 Cavitation

The thermodynamics for homogeneous and heterogeneous nucleation can be provided by CNT, as briefly described by the following sections. More information on the same has been provided in the appendix of this article.

#### 2.3.1. Homogeneous cavitation

When the (nano) porous media is dried, intrinsic capillary pressure in the order of megapascals is induced in the pore fluid. As this pressure is increased, intermolecular forces dominate over short-range repulsion forces and the liquid attains metastability. This metastability is ended by the homogeneous nucleation process or the spinodal breakdown of the fluid. The strong cohesion of water molecules enables water fluid to resist very high negative pressures. The stress for homogeneous cavitation in water has been predicted to be 140 MPa at 25 °C [22] and 200 MPa at 35 °C by the spinodal breakdown theory by Speedy.

The rate of homogenous nucleation of vapour bubbles $E_b$ [23–25] is given by:

$$E_b = \frac{N_A kT}{v_m h} exp\left[-\left(Q + \frac{16\pi\sigma_{lv}^3}{3(P)^2}\right)/kT\right] \qquad (5)$$

$N_A$ : Avagadro's number

$k$ : Boltzmann constant

$h$ : Planck's constant

$Q$ : activation energy required for molecular transport across the liquid-vapour interface.

For a measurable nucleation rate $E_b$ of 1 cm$^{-3}$s$^{-2}$ the cavitation stress could be approximated by[25]:

$$P^0 \approx \sqrt{\frac{16\pi\sigma_{lv}^3}{3kT \ln(\frac{N_A kT}{v_m h})}} \qquad (6)$$

Additionally, the energy cost of bubble creation $W_{homo}$ can be represented as the sum of interfacial energy required to create the bubble of radius $r$ and the work of negative pressure over the bubble area and ca be given by equation(7) [26] :

$$W_{homo} = \frac{4}{3}\pi r^3 \Delta g_v + 4\pi r^2 \sigma_{lv} \qquad (7)$$

$\Delta g_v$ : Difference in free energy per unit volume between the thermodynamic phase nucleation is occurring in and the phase that is nucleating and can be considered equal to $P$ as $P \gg p_{sat}$.

Furthermore, at sufficiently small $r$, the surface free energy term (second term on the right) would dominate, leading to initial positive values for $W_{homo}$ and attaining maximum value[27] at a critical radius $r^*$ which can be given as

$$r^* = 2\sigma_{lv} / \frac{RT}{\rho_l} \ln\left(\frac{p_v}{p_{sat}}\right) \tag{8}$$

Equation 8 suggests that cavitation would take place if the size of the bubble is great than $r^*$.

Further, CNT assumes that the decay of the metastable liquid under tension proceeds via the formation of a small vapour bubble, whose growth is initially opposed by a free energy barrier. The energy barrier corresponding to the nucleation of a spherical vapour bubble of radius $r$, is given by

$$\Delta G^* = (16\pi\sigma_{lv}^3)/3(P)^2 \tag{9}$$

Equations (6), (7) and (9) suggest that as $\sigma_{lv}$ for fixed temperature decreases, the cavitation stress $P^0$, the energy cost of bubble creation $W_{homo}$ as well as the energy barrier $\Delta G^*$, decreases. Therefore, modifying the fluid surface tension should modify the cavitation pressure.

### 2.3.2 Heterogeneous cavitation

The fundamental difference in the homogeneous and heterogeneous cavitation is the energy barrier of nucleation which is lower for heterogeneous nucleation (see appendix A2). The nucleation mechanism corresponding to heterogeneous cavitation is less understood but is governed by factors such as surface imperfections, wettability contrasts, among others, which are abundant in cementitious materials. Nevertheless, the stress of ~ 20 -25 MPa has been associated with heterogeneous cavitation of water in a variety of porous materials [28,29].

## 3. Materials and methods

As C-S-H is the continuous phase which dominates the pore structure of cementitious materials, samples were prepared with the aim to maximize the amount of C-S-H and minimise the amount of other phases. To achieve this, white cement clinker was used for its high content of calcium silicate phases. To further produce C-S-H, silica fume was added to react with calcium hydroxide.

White cement clinker provided by Aalborg Portland was used in this study. The composition of the anhydrous powder was analysed using XRD Rietveld refinement and was found as $C_3S$ (65.3 wt.%), $C_2S$ (28.8 wt.%), $C_3A$ (1.9 wt.%) and CH (2.51 wt.%) with the rest of the phases below 1%. To maximise hydration and minimise remaining anhydrous grains, the clinker was ground to $d_{50}$ and $d_{90}$ values of ~ 6 and 18 µm respectively. Condensed silica fume (SF) was used in this study which was provided by ELKEM. It was composed of 96% $SiO_2$ and had $d_{50}$ ~ 0.478 µm. Table 1 shows the XRF analysis of the white cement clinker and SF used in this study.

*Table 1 -Values are given in weight %*

| Chemical content | $SiO_2$ | $Al_2O_3$ | $TiO_2$ | MnO | $Fe_2O_3$ | CaO | MgO | $K_2O$ | $Na_2O$ | $SO_3$ | $P_2O_5$ | total XRF - 950 °C |
|---|---|---|---|---|---|---|---|---|---|---|---|---|
| Clinker | 25.25 | 2.07 | 0.05 | 0.01 | 0.13 | 69.91 | 0.64 | 0.15 | 0.22 | 0.19 | 0.14 | 99.48 |
| SF | 96.19 | 0.19 | 0 | 0.01 | 0 | 0.23 | 0.45 | 0.8 | 0.16 | 0 | 0.07 | 99.1 |

The cement and paste compositions were optimized throughout multiple iterations to maximize the C-S-H content of the hydrated paste sample. The following procedure has been chosen: ground white-cement clinker (84.5% by wt.); hemihydrate 5.5% by wt. and silica fume 10% by wt. were weighed and homogenised. Thereafter, this homogenised binder was mixed with water to binder (w/b) ratio of 0.7 using high-speed mechanical mixer Ultraturrax at 24000 rpm initially for 90 s and followed by another 30 s during the lowering of speed (24000-0 rpm). Then, the sample was cured underwater at 55 °C for 90 days to accelerate the hydration. This sample is referred to as the 'control sample' in this article. Two other samples were produced, incorporating a liquid shrinkage reducing admixture (SRA). The SRA used was provided by SIKA (SIKA control 40) and added in the amount of 3% and 2% by weight, replacing water. Liquid SRA contained 0.2 % solid. The samples with SRA were prepared following the same procedure as the control sample and were also cured at 55 °C for 90 days.

For X-ray diffraction analysis and Rietveld refinement, fresh discs of 3 mm thickness were cut from the sealed cylinders after 90 days. A Bruker D8 Advance diffractometer with monochromatic Cu-Kα radiation (λ = 1.541 Å) and a step-size of approximately 0.02° was used to acquire data in the range of 5° to 70° (2θ). The degree of hydration of clinker phases was calculated by Rietveld refinement using Topas software, with quartz as the external standard following the procedure described in detail elsewhere[30].

Further characterizations were made on samples that were dried using the solvent exchange method. This method was selected as it is considered to be the least damaging among other drying techniques [31–33]. The solvent exchange was carried out in the following manner: a disc of thickness ~ 2.5-3 mm was cut from the hardened cement cylinders and stored in isopropanol, which was replaced after 1 hour and 1 day. After 3 days of isopropanol immersion, the samples were stored in desiccators with silica gel beads under vacuum conditions. Thereafter, this disc was crushed and 30 mg (± 2) of the resulting powder was used for thermogravimetric analysis (TGA) (Netzsch STA 449 F3''Jupiter") from 20 to 700 °C in a nitrogen atmosphere with a heating rate of 10 °C/min. The portlandite content was calculated from the TGA analysis with the tangent method.

Scanning electron microscope (SEM) images were acquired using the FEI Quanta 200 microscope coupled with a Bruker XFlash 4030 EDS detector. A 2 mm thick disc (solvent exchanged) was impregnated with epoxy resin in a mould under vacuum. The samples were subsequently polished down to 1 μm using a diamond spray and a petrol-based lubricant before analysis.

For porosity analysis of the control sample, mercury intrusion porosimetry was carried out on crushed samples that were investigated up to a maximum pressure of 400 MPa using a Pascal 140/440 Porosimeter (Thermo Scientific). The surface tension of the mercury was considered as 0.48 N/m.

$^1$H NMR Carr–Purcell–Meiboom–Gill (CPMG) measurements were made for porosity analysis of the control sample. The crushed sample was equilibrated at 0.95 $p_v/p_{sat}$ for 6 months. The experiments with 256 echoes logarithmically spaced between 60 μs and 0.5 s were made on a bench top magnet operating at 20 MHz using Kea$^2$ spectrometer (Magritek, New Zealand). The 90° excitation pulse length was 6 μs and the spectrometer dead time was 10 μs. 8 points per echo were acquired with a dwell time of 1 μs. The experimental repetition time ($\tau_{rd}$) was 1 s and the number of scans was 1024.

Sorption isotherms were acquired using dynamic vapour sorption (DVS). Measurements were carried out on as-cured saturated samples using DVS Adventure (Surface Measurement Systems). The samples were cured for 90 days for this analysis. The sorption temperature was 20°C for all samples. The control sample was measured at 50 °C, additionally. Samples weighing ~25 mg were crushed and subjected to drying (or desorption) to have a starting point at 0.98 $p_v/p_{sat}$, followed by step-wise data acquisition from 0.98 $p_v/p_{sat}$ to 0.03 $p_v/p_{sat}$ with a step size of 0.5. For control sample data was acquired at additional steps, as well. Mass equilibrium was identified as the point at which the mass change was less than 0.0001mg/min or exceeding the assigned time limit which was 1440 mins for the control sample. For all the other samples (including the control sample measured at 50 °C) the time limit was 1440 mins for 0.98 $p_v/p_{sat}$ step and 480 mins at all other steps. If the equilibrium was not attained in the given time interval, the moisture content was extrapolated by using the following asymptotic function for moisture ratio $MR$ as:

$$MR = \frac{m_t - m_0}{m_{eq} - m_0} = 1 - exp(-k_m t) \qquad (10)$$

$m_t$ : mass at time $t$

$m_0$: mass at time 0

$m_{eq}$: Equilibrium mass at a given $p_v/p_{sat}$

$k_m$: fitting parameter

This equation has been previously used in several investigations on moisture sorption and is based on the analytical solution of the diffusion equation[11,34–36].

## 4. Results and analysis

### 4.1 Analysis of hydrated cement pastes

Phase assemblage of the hydrated cement pastes for the control sample and the sample prepared with 3 % SRA were determined after 90 days of hydration at 55°C using Rietveld refinement. Portlandite content was additionally measured by TGA. Table 2 provides information on the phase assemblage of the samples. The data were normalised per 100g of paste.

*Table 2 – Phase assemblage of the pastes determined using XRD Rietveld refinement and portlandite content determined using TGA. Measurement error is ± 2%*

| Phase composition | | Control sample | Sample with 3 % SRA |
|---|---|---|---|
| Alite ($C_3S$) | | 3.4 | 5.6 |
| Belite ($C_2S$) | | 4.3 | 3.9 |
| Ettringite (AFt) | | 3.6 | 5.5 |
| Portlandite (CH) | Rietveld | 8.1 | 7.6 |
| | TGA | 8.9 | 10.6 |
| Amorphous content | | 80.5 | 77.2 |

The proportion of the remaining clinker phases and hydrated compounds were found to be statistically equivalent between the control sample and the sample prepared with the SRA. The degree of hydration of clinker phases was ~ 90 % after 90 days of curing at 55°C for the two samples studied.

Figures 1 (a) and (b) show the SEM micrographs for the control sample and the sample prepared with 3 % SRA, respectively. The objective of the SEM imaging was to ensure that the use of 3 % SRA did not lead to agglomeration phenomena (for example of the silica fume) and that the sample was homogeneous. No distinguishable difference was observed between the microstructure of both samples. EDS analysis indicated that the Ca/Si ratio of the C-S-H was ~ 1.4-1.5 for both samples.

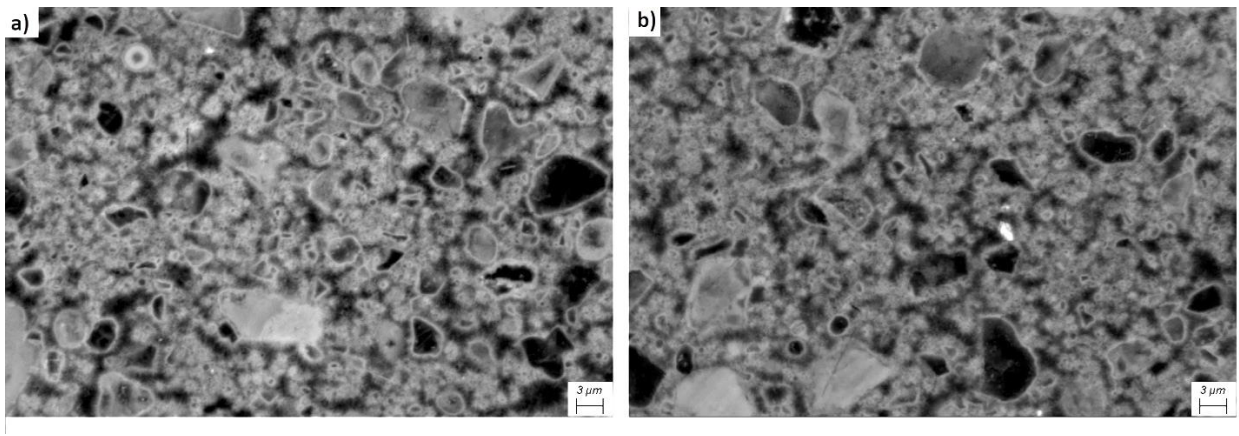

*Figure 1: SEM micrographs for (a) control sample and (b) sample containing 3 % SRA.*

Porosity analysis of the control sample was carried out using MIP and $^1$H NMR. Figure 2 (a) shows the intruded pore volume as the function of applied pressure, using MIP. The Washburn equation was used to calculate the equivalent pore entry radii. Considering the contact angle of 140°, the critical pore radius (which is defined as the pore size where the steepest slope of the cumulative intrusion curve is recorded) was estimated to be 20 nm. Additionally, threshold pore entry radius (which is interpreted as the minimum radius that is geometrically continuous throughout the whole sample) was found to be 33 nm. An additional scale considering a contact angle of 120° for determination of pore radius is presented in this figure.

Figure 2 (b) shows the multi-exponential fitting applied to analyse the $^1$H NMR CPMG data to separate the mobile water signal into different components. The fitting of NMR data was done by minimizing the squared error of fit. In this analysis, the ratio of the relaxations times was fixed such that $T_2$ value of gel pore water was equal to 3 times $T_2$ value of interlayer water. The shortest time and the different amplitudes were allowed to vary, as applied in several previous studies [37,38]. $T_2$ values of 192 and 576 µs were attributed to the interlayer and gel pore water of the C-S-H with a fraction of 26.6 and 56.2% of the total signal, respectively. This leads to a gel to interlayer amplitude ratio of 2.13, which is comparable to several other studies for mature white cement pastes[39][40].

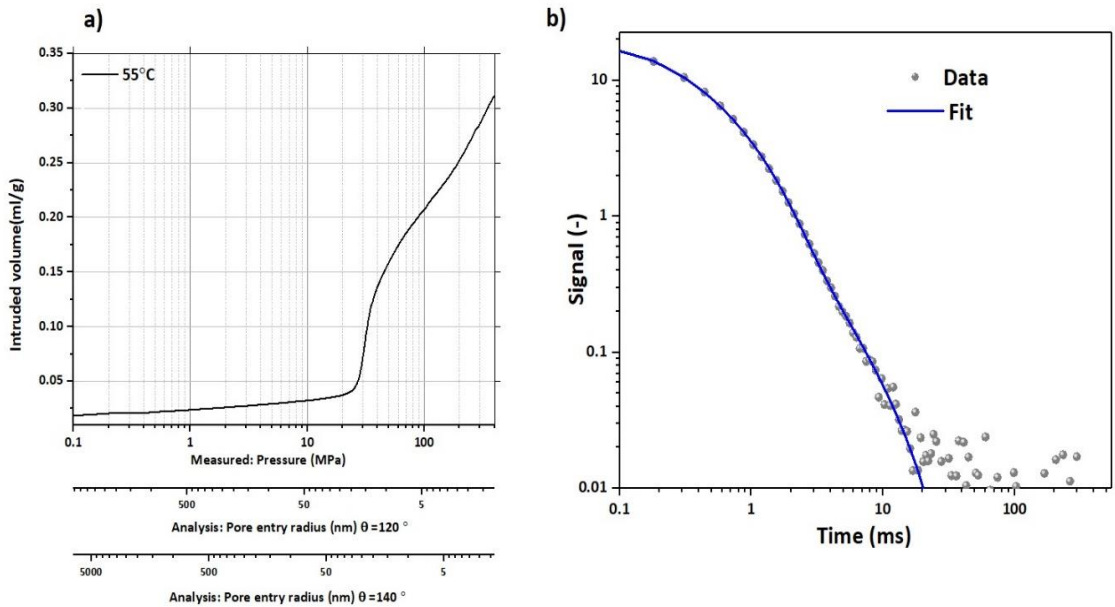

*Figure 2: Cumulative pore volume as obtained from the MIP analysis using 140° as a contact angle (additional scale corresponding to pore entry radius with contact angle 120 ° has also been plotted) and (b) CPMG decay data as obtained from $^1$H NMR relaxometry for the control sample and the corresponding multi-exponential fit.*

4.2 Sorption isotherms results

Figure 3 shows the desorption isotherm results of the control sample and the samples prepared with 2 and 3% SRA (measured at 20°C) and for the control sample measured at 50°C. For the control sample, two main desorption steps can be observed, one around 0.83 $p_v/p_{sat}$ and another ~ 0.3 $p_v/p_{sat}$ (indicated by the arrows in Figure 3). For the samples prepared with 3 % SRA, these steps were shifted to ~ 0.88 and 0.48 $p_v/p_{sat}$ . For the samples prepared with 2 % SRA, the steps were found to be shifted to 0.88 and 0.38 $p_v/p_{sat}$ . Further, for the control sample measured at 50 °C the step at ~ 0.43 $p_v/p_{sat}$ was observed.

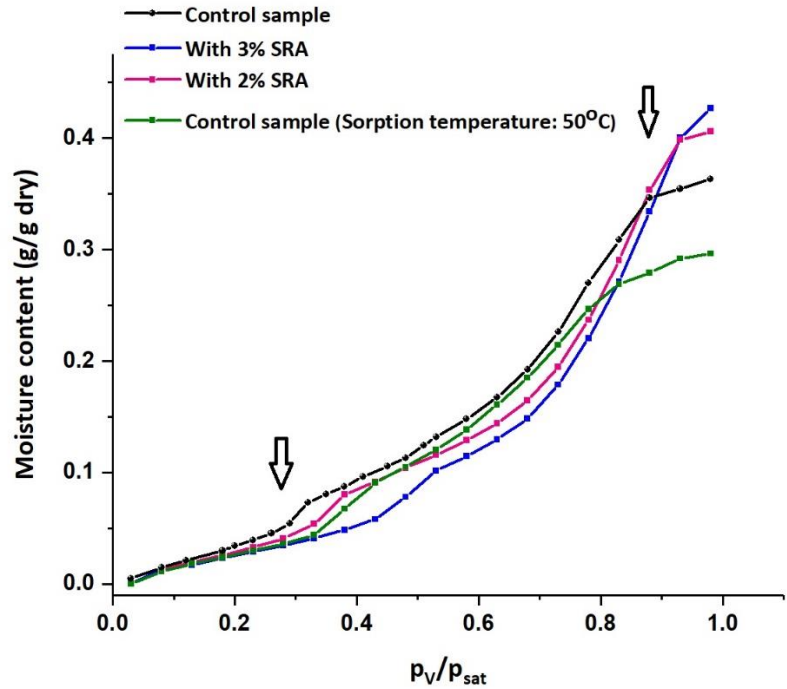

*Figure 3: Desorption isotherms for the control sample and the samples prepared with 2 and 3 % SRA at 20 °C and desorption isotherm for the control sample acquired at higher sorption temperature of 50 °C.*

### 4.2.1 Detailed analysis: Moisture loss, equilibration time and intrinsic capillary pressure

Figures 4 (a) and (b) show the moisture loss (%) as the function of equilibration time ( time required to reach dm/dt = 0.0001 mg /min) for each partial pressure step for the control sample and the sample prepared with 3 % SRA in the range 0.03-0.5 $p_v/p_{sat}$, respectively. For the control sample, the assigned time limit of 1440 mins was found to be sufficient for all the steps to reach equilibrium. For all the other samples, when the specified dm/dt was not reached for a given step, the final moisture content and equilibration time was determined by extrapolation using equation 10. Higher moisture loss was observed at around 0.32-0.29 and 0.48-0.43 $p_v/p_{sat}$ for the control sample and the sample prepared with 3 % SRA, respectively as compared to other steps. For the sample prepared with 2 % SRA  and for the control sample measured at 50°C more moisture loss was observed at around 0.38-0.33 and 0.43-0.48 $p_v/p_{sat}$, respectively (not shown here).

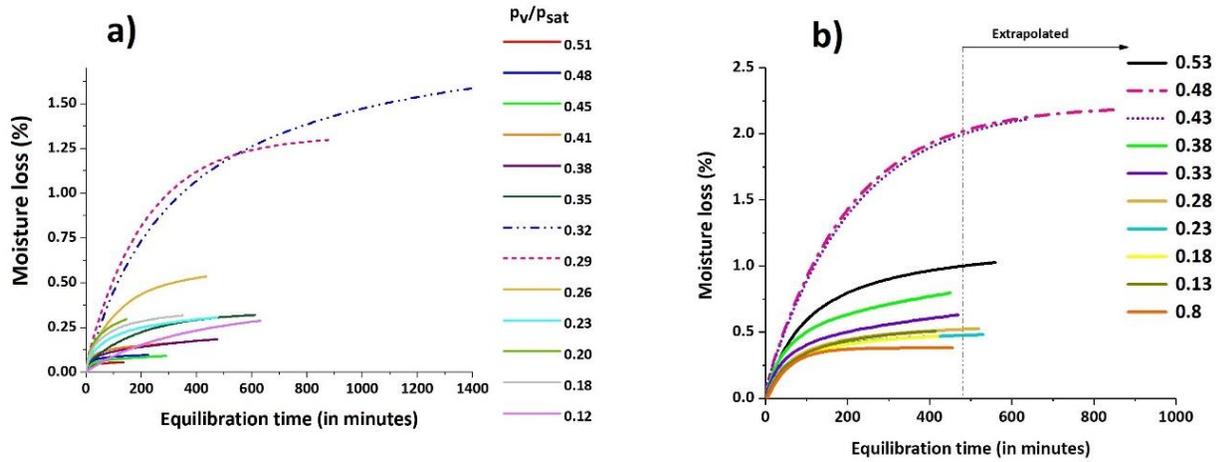

*Figure 4: Moisture loss as the function of equilibration time measured at 20°C for (a) the control sample and (b) the sample prepared with 3 % SRA, in desorption regime of 0.03-0.5 $p_v/p_{sat}$. Data with high moisture losses are indicated with a dotted line.*

Figures 5 (a) and (b) show the moisture loss (%) as the function of equilibration time for each partial pressure step for the control sample and the sample prepared with 3 % SRA in the range of 0.58-0.93 $p_v/p_{sat}$. More moisture loss was observed at around 0.83-0.78 and 0.88-0.83 $p_v/p_{sat}$ step for the control sample and the sample prepared with 3% SRA, respectively as compared to other steps. The sample prepared with 2 % SRA also showed higher moisture loss around 0.88 $p_v/p_{sat}$ step (not shown here). Though, no such distinct behaviour was observed for the control sample measured at 50°C in this regime. From the analysis shown in figures 4 and 5, it was observed that the steps that had more moisture loss also had higher equilibration time.

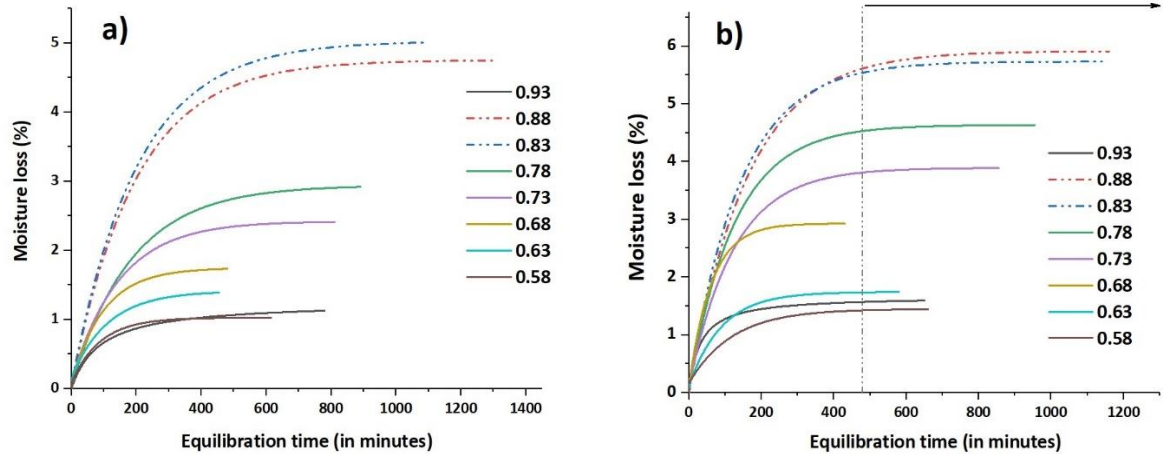

*Figure 5: Moisture loss as the function of equilibration time for (a) control sample and (b) sample prepared with 3 % SRA, in regime of 0.58-0.93 $p_v/p_{sat}$.*

From equation 1, the induced capillary pressure can be calculated as the function of $p_v/p_{sat}$ (for instance, 0.83 $p_v/p_{sat}$ would induce capillary stress of 25 MPa). Figure 6 (a) and (b) show the equilibration time as a function of the stress-induced in the pore fluid for the control sample and sample prepared with 3 % SRA. It was observed that induced stress of ~ 20 MPa and ~ 155 MPa results in higher moisture loss and equilibration time for the control sample (figure 6 (a)). For the sample containing 3 % SRA, the induced stress of ~ 16 MPa and ~ 101 MPa results in higher moisture loss and equilibration time (figure 6 (b)). Further, induced stress of ~ 17 MPa and ~ 130

MPa leads to higher moisture loss and equilibration time for the sample containing 2 % SRA and induced stress of ~ 110 MPa results in higher moisture loss and equilibration time for the control sample measured at 50°C (not shown here).

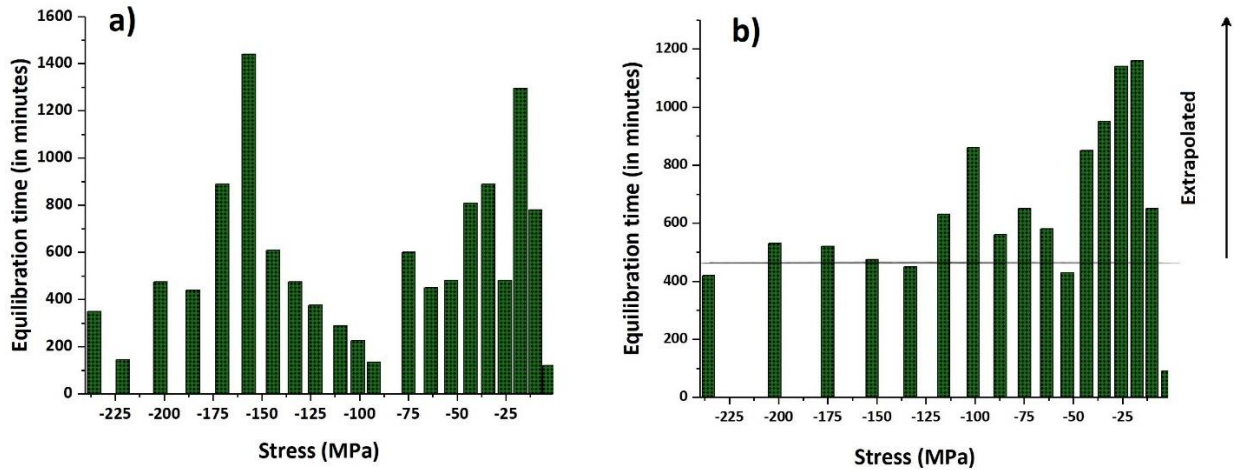

*Figure 6: Equilibration time as the function of induced stress for (a) control sample and (b) sample containing 3 % SRA.*

### 4.3 BJH analysis

Figure 7 shows the BJH pore size distribution obtained from the desorption isotherms presented in Figure 3. The surface tension and density of water were corrected when changing the measurement temperature and for the presence of admixture for the BJH analysis. A reduction of 15% in the surface tension of water (61.93 mN.m$^{-1}$ ) and 5% in density of water (948 kg.m$^{-3}$) has been considered for the control sample measured at 50 °C following the observations and conclusions of previous studies [41–43]. For the sample prepared with 2 and 3 % SRA, reduced values of surface tension (57 mN.m$^{-1}$ for 3 % SRA) and (65 mN.m$^{-1}$ for 2 % SRA) was used following the experimental results in previous studies [44] [45].

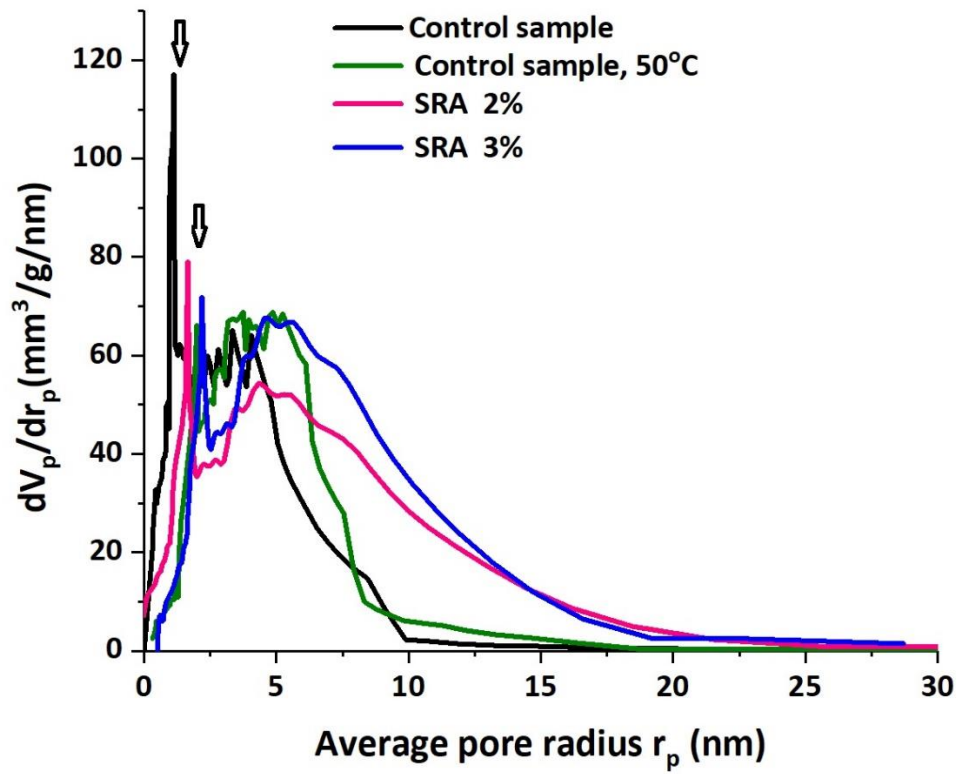

*Figure 7: Pore size distribution of the control sample measured at 20°C and 50°C and sample containing 3 % SRA.*

As commonly observed, a large population of pores is located at around $r_P$ ~1.7 nm in the pore size distribution of the control sample (measured at 20°C). The corresponding sharp peak was observed at $r_P \sim$ 1.9 nm, 2.0 nm and 2.2nm for the sample containing 2 % SRA, control sample measured at 50°C and the sample containing 3 % SRA, respectively.

# 5. Discussion

## 5.1 Heterogeneous and homogeneous cavitation

The steep desorption step observed at around 0.8 $p_v/p_{sat}$ for most samples (except for the sample measured at 50°C) in the desorption isotherms can be attributed to drying by heterogeneous cavitation. Cavitation under capillary pressure of 20-25 MPa would be governed by the population of capillary pores greater than $r^* \approx 11.2$nm ((section 2.3.2 and equation 8). The absence of heterogeneous cavitation in the desorption isotherm of the control sample measured at 50°C could be since a majority of pores bigger than this size could already have dried out at higher $p_v/p_{sat}$ because of exposure to 50 °C environment.

As the drying continues, intrinsic capillary pressure of ~ 150 MPa would be induced in the pore fluid around 0.3 $p_v/p_{sat}$. Due to this high tensile stress, the metastable pore fluid would be changed into vapor phase either by the spinodal breakdown or by bubble nucleation and subsequent homogeneous cavitation. As predicted by classical nucleation theory, reducing $\sigma_{lv}$ decreases the work of nucleation $W_{homo}$ and energy barrier associated with the creation of bubble $\Delta G^*$ (equations 8 and 9) which in turn, reduces the magnitude of intrinsic capillary stress that water could sustain before homogeneous cavitation. This explains why reducing surface tension at the liquid-vapor interface shifts the desorption step commonly observed around 0.3 $p_v/p_{sat}$ to higher $p_v/p_{sat}$ for the samples with reduced surface tension. For instance, the step with the starting point around 0.3 $p_v/p_{sat}$ for the control sample shifts to 0.48 $p_v/p_{sat}$ for the sample prepared with 3 % SRA as the homogeneous cavitation pressure reduces from ~ 150 MPa for the control to ~ 110 MPa (equation 1). Further, homogeneous cavitation, therefore, more likely explain the desorption step at 0.3 $p_v/p_{sat}$ than spinodal which is not affected by the nucleation barrier.

Further, the overall equilibration time is high for the steps where evaporation induced cavitation takes place as shown by figures 4 and 5. We propose the following mechanism for higher equilibration times corresponding to evaporation induced by (homogeneous and heterogeneous) cavitation: when a cavitation event occurs close to a drying front, the fluid would evaporate. Otherwise, the trapped vapor would cause re-filling of the pores in the vicinity. The cavitation events would stop only when the nucleation sites cease to exist, moisture transport stops, and intrinsic capillary pressure is lowered, which therefore extends the overall equilibration time.

To have first heterogeneous cavitation and then homogeneous cavitation, the pores in which homogeneous cavitation occurs must be isolated from the rest of the porosity. Additionally, from equation 8, for homogeneous cavitation to be energetically favourable at an induced tension of ~ 150 MPa (or $0.3 p_v/p_{sat}$ ) , a bubble must be larger than $r^* \approx 1\ nm$. This indicates that such events are only feasible in the gel pores of the C-S-H that are completely isolated from the pores where heterogeneous cavitation events can take place. Comparison of the pore distribution from other techniques and moisture loss content from the desorption isotherm could provide an insight into this hypothesis.

## 5.2 Comparative overview of porosity and DVS data

Porosity analysis of the control sample was carried out by two independent methods.

The first method was based on evaluating the content of C-S-H interlayer water by using phase assemblage data from Rietveld refinement and making a calcium mass balance. The mass of calcium within the C-S-H was calculated taking the calcium provided by the reacted clinker phases and subtracting the calcium present in Ettringite and Portlandite. A ratio of $H_2O$/Ca of 0.482 from [39] was used to calculate how much water is present in the C-S-H interlayer. Subsequently, a gel/interlayer ratio of 2.13 as measured by [1]H NMR CPMG analysis was used to calculate the amount of gel water in the sample. The capillary porosity was obtained by subtracting the sum of interlayer, gel and chemically bound water from the total porosity.

In the second method, multiple techniques were used for the porosity analysis of the control samples: ignition test, careful weighing before and after the solvent exchange, MIP and TGA analysis. The effective water to binder ratio was found to be 0.78 after igniting the sample to 1050°C and accounting for the loss of carbonates and the loss on ignition (LOI) from XRF analysis. It was assumed that MIP can intrude all the accessible voids, capillary porosity and a part of gel porosity. TGA analysis was used to determine chemically bound water content.

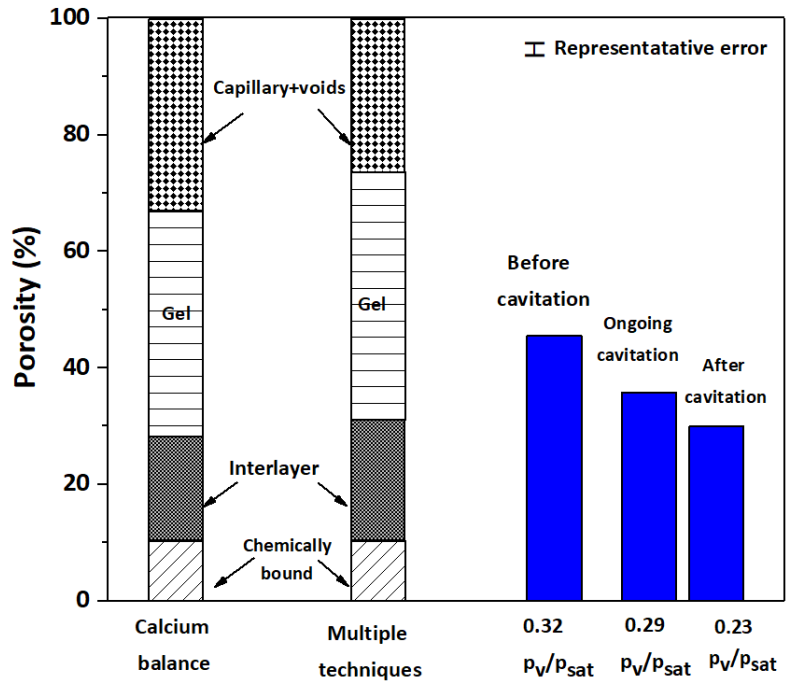

Figure 9: Water content overview of the control sample obtained using two independent methods. The blue bars correspond to the amount of water at 0.32 0.29 and 0.23 $p_v/p_{sat}$.

Figure 8 provides a comparative overview of the repartition of water obtained from both porosity analysis methods and the amount of water present in the sample before, during and after cavitation (shown in blue), as obtained from the sorption data using DVS. From this comparative analysis, it was found that at least 26 % of the gel pores were isolated from the rest of porosity by C-S-H interlayers and were emptied homogeneous cavitation.

Figure 9 shows a schematic representation of the homogeneous cavitation events as we perceive it. (i) C-S-H is saturated and the black arrows represent the drying front (ii) as the $p_v/p_{sat}$ reduces, drying occurs by receding meniscus (iii) when $p_v/p_{sat}$ reduces to 0.3, the tensile stress in the metastable pore fluid of interlayer space would reach ~ 150 MPa, the meniscus cannot recede due to the small radius of the interlayer. However, the fluid in the interlayer spaces and the gel pores behind, experience the tensile pull which is equal to homogeneous cavitation stress. The existing bubbles would also experience this tensile pull due to which they can expand in the gel pore and result in cavitation (iv) water in the gel pore and interlayer is changed into water vapor can leave the sample through drying front.

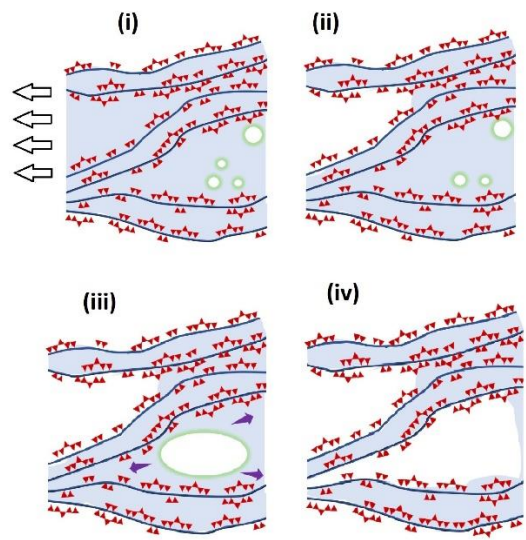

*Figure 9: A feasible homogeneous cavitation event in C-S-H, black arrows represent the drying front, (i) saturated C-S-H (ii) drying occurs by retreating meniscus (iii) Bubble expansion in the gel pore due to induced tensile pull (iv) drying by cavitation.*

### 5.3 Comparison with grey cement pastes

It has been observed that for white cement pastes and high w/b, the (MIP) 'accessible' gel porosity increases[46]. Müller *et al*. investigated several white cement pastes with varying w/b from 0.32 to 0.48 after 28 days of hydration using MIP and concluded that mercury intruded small volumes of C-S-H gel porosity at high mercury pressures. For instance, for the sample with 0.48 w/b ratio, the gel pore content was calculated to be 30. 3 %, out of which 8. 5% of the pore volume could be intruded by mercury, whereas in grey cement with w/b of 0.4, gel pores could not be intruded[47]. In this study, the total gel porosity was found to be 42.27 % for the control sample using multiple techniques, out of which 14.56 % volume could be intruded by MIP. The fact that the part of the gel pores is isolated from the rest of the porosity could also indicate that this porosity lies in the inner product rather than the outer product of C-S-H.

### 5.4 Pore size distribution from desorption isotherms

Referring to figure 7, starting with the same sample, it is very unlikely that desorption temperature of 50°C can introduce the microstructural changes which would create a large population of pores around $r_P$ ~ 1.9 nm as compared to existing 1.7 nm for the control sample when measured at 20°C. Therefore, due to the anomalies of water, it is not recommended to use desorption isotherm of cementitious materials for pore size distribution. Further, due to many underlying assumptions, using classical methods such as BJH analysis could result in very erroneous conclusions.

## 6. Conclusions

In this paper, we explain the role of cavitation in drying cementitious materials using classical nucleation theory. The change in cavitation pressure with the addition of SRA and increase in temperature support the homogeneous cavitation over spinodal. The presence of pre-existing bubbles, surface imperfections, or surface wettability contrasts significantly reduces the cavitation pressure in cementitious materials and promotes heterogeneous cavitation first. Thereafter, the homogeneous cavitation occurs in the part of the gel porosity which is isolated from the rest of the porosity. Cavitation results in higher moisture loss and longer overall equilibration time. The results presented in this study also provides insights into the limitations of investigating microstructure using desorption isotherm and subsequent erroneous conclusions on the pore size distribution using classical methods such as BJH theory.

# Appendix

## A1 Phase diagram of water

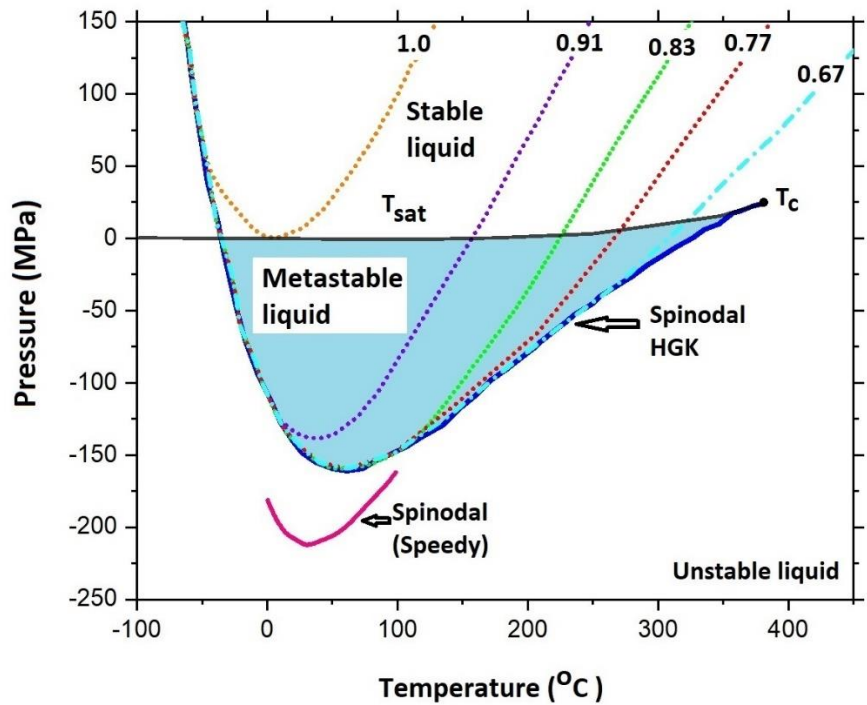

*Figure A1: Phase diagram of water based on the extrapolation of HGK equation of state of water and Speedy's equation for different densities. Reproduced with permissions.*

Figure A1 shows the phase diagram of water. The isochores presented in figure A1 are acquired by extrapolation of the HGK (Haar-Gallagher-Kerr H-G-K) equation below the equilibrium vapour pressure line[22] as the function of the density of water.

The H-G-K equation itself is based on high precision experimental data collected in the positive pressures regime[48]. As shown in figure A1, if the tensile stress on the liquid is increased intermolecular forces dominate over short-range repulsion forces and the liquid attains a state of metastability. This metastability is ended by the homogeneous nucleation process. Spinodal decomposition of water is feasible in the unstable region of the phase diagram of water as shown in figure A1. The phase separation due to spinodal decomposition occurs without the nucleation barrier. It is also important to note that the density of the water in pore also influences the cavitation pressure. As the density of water in confinement is different from the bulk and vary from one sample to the other, the tensile stress sustained before cavitation can also vary up to a certain extent. Additionally, using Tolman type corrections on the magnitude of surface tension can also result in lowering the overall magnitude of tensile stress.

## A2 Heterogeneous cavitation

Heterogeneous nucleation is dependent on the surface chemistry of the system, and bubble nucleates on the pore wall on the available nucleation site. This suggests that nucleation barrier or work for heterogeneous cavitation is lower as compared to homogeneous cavitation where a nucleation site is absent. Additionally, the work of heterogeneous nucleation is determined by the differences in the surface areas $\Delta S$ and volumes of the unstable and stable bubbles $\Delta V$ and is given as[49].

$$W_{hetro} = \gamma(\Delta S) - \frac{2\Upsilon}{r}(\Delta V) \qquad (10)$$

Comparing the work for homogeneous nucleation as given by equation (7) and equation (9), it can be concluded that $W_{hetro} < W_{homo}$.

## A3 Results of cavitation in this study vs. direct observations of cavitation in nanofluidic devices

Understanding the nucleation mechanism and mobility of the bubble is not within the scope of this study. However, some relevant results from direct observations of cavitation events in transparent nanofluidic devices can provide some insights that could be helpful for future studies in this field.

Duan *et al.* fabricated silica nanochannels with the varying height from 20 nm-120nm and observed the role of cavitation in the drying of the nanochannels [17]. The authors also observed faster drying or evaporation rate that was an order of magnitude higher than what is obtained by receding meniscus which is comparable to the results presented in this study where high moisture loss was associated with cavitation events. The evaporation induced cavitation was found to be aided by advective liquid transport and the bubbles showed unusual motion as well as transitional stability which was attributed to balance between two competing mass fluxes driven by thermocapillarity and evaporation.

Vincent *et al.* experimentally investigated a synthetic sample with extreme ink-bottle geometry that had nanometers necks connecting to micrometres voids. The drying dynamics were analysed by time-lapse photography. The drying stress was noted to be in the order of 24-30 MPa and the deterministic mass transport and the stochastic events were suggested to govern heterogeneous cavitation[50]. The authors also observed that the cavitation events were interspersed and distributed over the entire timespan until all the voids were empty. These observations can also explain why in the porous media with interconnected pores, cavitation would take longer equilibration times than usual drying by receding meniscus.

## Acknowledgements

The research leading to these results has received funding from H2020-MSCA-ITN ERICA project with grant agreement ID 764691. Authors are grateful to Prof. Peter J. McDonald, Dr Agata M. Gajewicz-Jaromin and Magdalena Janota from the University of Surrey for $^1$H NMR relaxometry data and analysis.